\documentclass[conference]{IEEEtran}

\usepackage{algorithmic}
\usepackage{amsmath,amssymb,amsfonts}
\usepackage{balance}
\usepackage{cite}
\usepackage{comment}
\usepackage{dirtree}
\usepackage{fontawesome}
\usepackage[T1]{fontenc}
\usepackage{graphicx}
\usepackage{hyperref}
\usepackage{multirow}
\usepackage{pgfplots}
\usepackage{pifont}
\usepackage{textcomp}
\usepackage{tikz}
\usetikzlibrary{patterns}
\pgfplotsset{compat=1.18}
\usepackage{xcolor}
\usepackage{xspace}
\usepackage{url} 
\usepackage{setspace}

\usepackage[acronym]{glossaries}
\newacronym{ann}{ANN}{Approximate Nearest Neighbor}
\newacronym{asr}{ASR}{Automatic Speech Recognition}
\newacronym{ir}{IR}{Information Retrieval}
\newacronym{nlp}{NLP}{Natural Language Processing}
\newacronym{rag}{RAG}{Retrieval Augmented Generation}
\newacronym[plural=LLMs,firstplural=Large Language Models(LLMs)]{llm}{LLM}{Large Language Model}

\def\BibTeX{{\rm B\kern-.05em{\sc i\kern-.025em b}\kern-.08em
    T\kern-.1667em\lower.7ex\hbox{E}\kern-.125emX}}

\newcommand{\frameworkname}[0]{SoccerRAG\xspace}

\definecolor{forestgreen}{RGB}{34,139,34}
\definecolor{darkyellow}{RGB}{255,223,0}
\definecolor{debianred}{rgb}{0.84, 0.04, 0.33}
\definecolor{midnightblue}{HTML}{006795}
\definecolor{royalpurple}{HTML}{613F99}

\newcommand{\mycircle}[1]{\tikz{\filldraw[draw=#1,fill=#1] (0,0) circle [radius=0.075cm];}} 
\newcommand{\mysquare}[1]{\tikz{\filldraw[draw=#1,fill=#1] (0,0)
rectangle (0.15cm,0.15cm);}}
\newcommand{\mytriangle}[1]{\tikz{\filldraw[draw=#1,fill=#1] (0,0) --
(0.2cm,0) -- (0.1cm,0.2cm);}}

\newcommand{\yes}[0]{\textcolor{forestgreen}{\ding{52}}}
\newcommand{\no}[0]{\textcolor{debianred}{\ding{54}}}

\newcommand{\high}[0]{\mycircle{forestgreen}\xspace}
\newcommand{\medium}[0]{\mysquare{darkyellow}\xspace}
\newcommand{\low}[0]{\mytriangle{debianred}\xspace}

\newcommand{\gptfour}[0]{GPT-4.0-Turbo\xspace}
\newcommand{\gptthree}[0]{GPT-3.5-Turbo\xspace}

\usepackage[linewidth=1pt]{mdframed}

\begin{document}
\bstctlcite{IEEEexample:BSTcontrol}

\title{\frameworkname: Multimodal Soccer Information Retrieval via Natural Queries}

\author{\IEEEauthorblockN{Aleksander Theo Strand}
\IEEEauthorblockA{\textit{OsloMet, TET Digital AS}\\
Oslo, Norway \\
0009-0008-2749-2347}
\and
\IEEEauthorblockN{Sushant Gautam}
\IEEEauthorblockA{\textit{OsloMet, SimulaMet} \\
Oslo, Norway \\
0000-0001-9232-2661}
\and
\IEEEauthorblockN{Cise Midoglu}
\IEEEauthorblockA{\textit{SimulaMet, Forzasys} \\
Oslo, Norway \\
0000-0003-0991-4418}
\and
\IEEEauthorblockN{Pål Halvorsen}
\IEEEauthorblockA{\textit{OsloMet, SimulaMet, Forzasys} \\
Oslo, Norway \\
0000-0003-2073-7029}
}

\maketitle

\begin{abstract}
The rapid evolution of digital sports media necessitates sophisticated information retrieval systems that can efficiently parse extensive multimodal datasets. 
In this paper, we introduce \frameworkname, an innovative framework designed to harness the power of \gls{rag} and \glspl{llm} to extract soccer-related information through natural language queries. By leveraging a multimodal dataset, \frameworkname supports dynamic querying and automatic data validation, enhancing user interaction and accessibility to sports archives. Our evaluations indicate that \frameworkname effectively handles complex queries, offering significant improvements over traditional retrieval systems in terms of accuracy and user engagement. The results underscore the potential of using \gls{rag} and \glspl{llm} in sports analytics, paving the way for future advancements in the accessibility and real-time processing of sports data.
\end{abstract}


\glsresetall
\section{Introduction}\label{section:introduction}

The rapid growth of digital sports content has created a demand for efficient retrieval systems that can understand and process natural language queries~\cite{gautam2022assisting,Gautam2023Oct}. The motivation behind our research is to address the challenges in retrieving specific content from extensive sports libraries using intuitive, natural language requests. This approach simplifies user interaction and enhances the accessibility of sports archives. Soccer, as one of the most popular sports worldwide, serves as a prominent use case and context for our proof-of-concept retrieval application. 

We propose \frameworkname, a framework for retrieving multimodal soccer information using natural language queries, from an augmented soccer dataset based on SoccerNet~\cite{SoccerNet,SoccerNetv2}, which includes game videos with image frames and audio, timestamped captions (transcribed audio), annotations for game events, and player information. In short: 

\begin{itemize}
    
    \item We highlight the opportunities presented by the rapid advancements in \glspl{llm} and \gls{rag}, and motivate the use of these technologies in the context of sports and more specifically for soccer analytics (Section~\ref{section:background}). 

    \item We propose a framework concept and design for the retrieval of multimodal soccer information through natural queries, called \frameworkname, which integrates components for data representation, feature extraction and validation, and database querying. Our conceptualization includes a proposed database schema for an augmented version of the SoccerNet dataset, as well as a novel extractor-validator chain (Section~\ref{section:framework}). 
    
    \item We provide an open source implementation for \frameworkname which is accessible under~\cite{SoccerRAG-GitHub} and fully reproducible through the instructions provided therein. 
    
    \item We present a preliminary evaluation of the \frameworkname framework through the analysis of the extractor-validator chain, query complexity, individual component contributions, and execution time, for different \glspl{llm} (Section~\ref{section:evaluation}), and discuss our insights (Section~\ref{section:discussion}).
    
\end{itemize}

\section{Background and Related Work}\label{section:background}

\glspl{llm} have revolutionized the field of \gls{nlp} and generative AI~\cite{Yang2023Jun}. The evolution of \glspl{llm}, from early rule-based models to contemporary GPT iterations, exemplifies significant technological advancements~\cite{Myers2024Feb}. The capabilities of these models extend beyond text generation to include reasoning, decision-making, and multimodal data fusion, addressing both the static nature of early models and the computational demands of training and fine-tuning~\cite{Xi2023Sep}.

\subsection{Multimodal Soccer Understanding}

Soccer, a globally beloved sport, serves as a rich domain for analytical exploration. The multifaceted aspects of soccer analytics are explored through the SoccerNet dataset and challenges~\cite{SoccerNet,SoccerNetv2}, a comprehensive repository of soccer broadcast recordings and various annotations, facilitating research in areas such as action spotting, video captioning, and game state reconstruction~\cite{Akan2023Jun}. Midoglu et al.~\cite{Midoglu2022Feb} focused on automatic event clipping, thumbnail selection, and game summarization through AI, to streamline the production of engaging soccer game highlights and summaries. There are interesting lines of research in soccer game understanding around event detection~\cite{Morra2020Jun,Rongved}, automatic game summarization~\cite{Gautam2022Oct}, highlight generation~\cite{Midoglu2024Feb}, caption generation~\cite{Mkhallati,Qi2023Oct}, and similar. Most of this research demonstrates that the use of multiple modalities (including but not limited to videos, images, text, audio, and structured metadata such as commentary text, event information, and team statistics) can enhance game understanding~\cite{gautam2022assisting,Gautam2023Oct}. 

\subsection{\acrfull{ir}}

\gls{ir} is focused on the extraction of relevant information from vast datasets based on user queries~\cite{chowdhury2010introduction}. It encompasses a range of techniques including indexing, search algorithms, and natural language processing, to organize and retrieve unstructured data efficiently~\cite{Boukhari2023May}. Recent advancements have leveraged machine learning and AI to enhance the precision of search outcomes through a better understanding of user intent and document semantics~\cite{Guo2020Nov}. Despite its progress, \gls{ir} continues to face challenges such as data volume management and privacy concerns, highlighting the need for innovative solutions in data retrieval~\cite{Olteanu2021Mar}.

\subsection{\acrfull{rag}}

\gls{rag} systems represent a paradigm shift in open-domain question answering, combining document retrieval with generative modeling to produce contextually relevant answers~\cite{Liu2022Jul, lewis2021}. The utility of vector databases in \gls{rag} systems, particularly in efficiently managing and retrieving high-dimensional vector representations of text, underscores their significance in enhancing the accuracy and relevance of generated responses~\cite{Yu2022Nov}. \gls{rag} has found application in different domains including enterprise contexts where \glspl{llm} are integrated with chatbots, enabling them to automatically derive more accurate answers from company documents and knowledge bases~\cite{Jeong2023Sep}.

\subsection{Terminology}

\textbf{\glspl{llm}} are advanced deep learning models capable of processing and generating human-like text~\cite{Zhao2023Mar}. Our research predominantly focuses on OpenAI's closed-source \gptthree~\cite{OpenAIGPT3} and \gptfour~\cite{OpenAIGPT4}, although we acknowledge the broader landscape of available \glspl{llm}. 
\textbf{Tokens} are the atomic elements used by \glspl{llm} for text generation and prediction. Tokens can represent varying levels of linguistic units, from single characters to multiple words, crucial for modeling and API cost calculations~\cite{Vaswani2017Dec}. 
Other concepts that are integral to \gls{llm} frameworks include \textbf{chains} which are sequences of tasks executed in a predefined order~\cite{Wu2022Apr}, and \textbf{agents} which are dynamic entities that use \glspl{llm} for decision-making and executing actions based on reasoning~\cite{Liu2023Aug}. 
\textbf{Vector databases} handle high-dimensional data, such as text and image embeddings~\cite{Ayesha2020Jul}, enabling quick similarity searches~\cite{Echihabi2021Jul} through \gls{ann} algorithms. They are crucial for integrating with \glspl{llm} across various industries, allowing for complex queries over extensive datasets, offering a faster alternative to traditional database search methods~\cite{Friedman2023May}. 
\textbf{Extractors} are specialized mechanisms that harness the capabilities of \glspl{llm} for efficient information extraction, and are traditionally manual, rule-based processes~\cite{Xu2023Dec}. They adapt to tasks via instructions or emulate JSON for structured output~\cite{Zhang2023May}. They offer tool calling for schema structuring, JSON mode for structured output, and prompting-based extraction for versatile retrieval, streamlining the costly traditional systems~\cite{LangChainExtractors}.

\subsection{Novelty}

The burgeoning interest in \gls{rag}, fueled by the rapid advancements in \glspl{llm}, has paved new avenues for exploring innovative use cases across various domains, notably within multimodal information retrieval~\cite{Cheng2024Jan}. Despite the wide applicability of \gls{rag} frameworks in enhancing the capabilities of generative AI for open-domain question answering and beyond, its potential in the sports domain, particularly within soccer analytics, remains largely untapped. Our research endeavors to bridge this gap by harnessing the multimodal datasets inherent to the soccer broadcast pipeline. By integrating \gls{rag} with multiple data modalities, we aim to develop a sophisticated system capable of understanding complex game scenarios and responding to user queries in a manner that resonates with the intricate language of soccer. The potential of our proposed system for video retrieval offers a promising avenue for users and broadcasters alike, enabling the recollection of game moments through natural queries. 

\section{Proposed Framework}\label{section:framework}

\begin{figure}[!b]
    \centering
    \includegraphics[width=1\columnwidth]{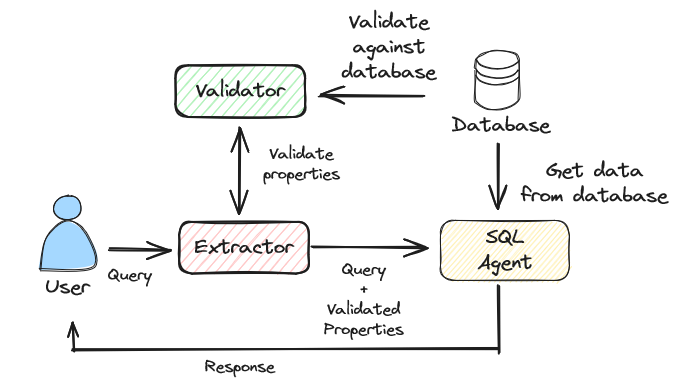}
    \caption{Overview of the proposed \frameworkname framework.}
    \label{fig:proposed-framework}
\end{figure}

Figure~\ref{fig:proposed-framework} presents an overview of our proposed \frameworkname framework, which includes four main components: \textbf{database}, \textbf{feature extractor}, \textbf{feature validator}, and \textbf{SQL agent}. The framework is intended for a scope where the dataset of interest is stored in a database with a predefined schema. The application flow is as follows: 
(1) The user provides a natural language query related to the contents of the database. 
(2) The user input is sent to an \gls{llm} along with the properties schema and a system prompt describing the properties the \gls{llm} should extract from the question. The \gls{llm} then returns a list of extracted properties relevant to the query. 
(3) Each extracted feature is checked against the appropriate table in the database using string matching algorithms. This step aims to correct spelling mistakes and abbreviations. Once a value is found, both the value and its primary key are added to the extracted value. 
(4) The cleaned user prompt is combined with system-specific prompts to guide the \gls{llm} in generating SQL queries that will answer the user's question. The constructed query is then passed to the SQL chain, which designs and executes the SQL queries against the underlying database. The SQL chain handles communication between the system and the database, retrieves the requested data, and prepares the results for presentation to the user. 
The complete \frameworkname framework can be replicated in an end-to-end manner by running our open source software, which is accessible under~\cite{SoccerRAG-GitHub}, through the instructions provided. A demonstration is also available~\cite{strand2024demo}.

\subsection{Database}\label{section:database}

The SoccerNet dataset consists of game broadcast videos and metadata information for 550 soccer games, from a number of top European leagues across multiple seasons. We have run \gls{asr} on all games with available audio using Open AI's Whisper~\cite{radford2022robust} to transcribe the audio, converting spoken commentary into textual data. Overall, for each game, we have files containing transcribed commentary (\textit{<1/2>\_half-ASR.json}), annotations describing key events during the game such goals, cards, fouls, etc. (\textit{Labels-v2.json}), and general game information such as home and away teams, lineups, score, referee, and other details (\textit{Labels-caption.json}). In the context of this work, an \textbf{event} refers to in-game events, \textbf{commentary} refers to the transcribed commentary from the game broadcast, and \textbf{caption} refers to a short description of an event. 

\begin{figure}[htbp]
    \scriptsize
    \centering
    \includegraphics[width=0.9\columnwidth]{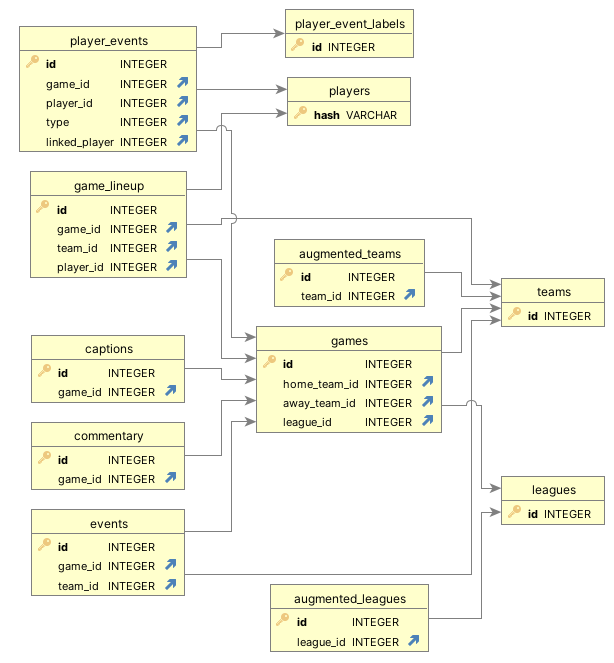}
    \caption{Schema for the structured SQLite database.}
    \label{fig:database-erd}
\end{figure}

To make this dataset usable for the SQL agent, we converted the JSON files into a structured SQLite database (detailed in Figure~\ref{fig:database-erd}) with a script utilizing SQLAlchemy~\cite{SoccerRAG-GitHub}. This script creates tables and columns based on a predefined schema, follows a systematic process of extracting league/season details, game details, lineups, events, commentary, and creates links between related entities. Clear and descriptive naming conventions for tables/columns were iteratively refined to improve the \gls{llm}'s understanding of the data structure. 

\subsection{Feature Extractor}

In the extractor chain, instructions are provided to the \gls{llm} to return a JSON structure, where pre-defined properties (defined in the \textit{schema.json} file under~\cite{SoccerRAG-GitHub}) are extracted from the given prompt. We use the Langchain framework for its easy adjustability to new tasks. The schema also includes a list of the tables and columns a property exists in in the database, as a preparation for the validation function.

\subsection{Feature Validator}

Extracted features can have spelling errors or include abbreviations, rendering them disparate from the representations stored in the database. This discrepancy necessitates a validation procedure. To address the challenge of abbreviations (e.g., the natural query referring to "ManU" in reference to Manchester United F.C.), we manually constructed auxiliary tables within the database and populated them with abbreviations pertaining to common soccer team and player nomenclature (\textit{augmented\_<teams/leagues>.csv} files under~\cite{SoccerRAG-GitHub}). These supplementary tables are also delineated in the extractor schema for properties that warrant their employment. 

The validation process commences by selecting the initial property key, which comprises a list of the extracted items for that property. If the property is associated with an auxiliary table, a search for a matching entry is performed within that table. In the event that no match is found, the system retrieves all items in the table linked to the property and computes the Levenshtein Distance~\cite{levenshtein1966binary} to ascertain the closest match for each item. Utilizing a predefined threshold for closeness, the system attempts to automatically select the optimal match, but may solicit user input for clarification in cases of ambiguity. 
Upon the validation of all items, the original prompt is returned with updated information containing the correct item designation, and where feasible, the corresponding foreign key in the database.

\subsection{SQL Agent}

The final component of the \frameworkname framework leverages the SQL agent from the LangChain framework~\cite{LangChainAgents}. To enhance the capabilities of the agent, we implemented a few-shot SQL \gls{rag} solution. By employing this approach, the system prompt for the agent is consistently accompanied by N number of queries that are pertinent to the user's question, where N is externally configurable. To identify the matching queries, we employ vector search facilitated by a FIASS database. This database is populated with data originating from a JSON file (\textit{sqls.json} under~\cite{SoccerRAG-GitHub}), which contains a collection of human crafted SQL queries. The vector search technique entails embedding both the user's question and the SQL queries into high-dimensional vector spaces, thereby enabling the identification of semantically similar queries through the calculation of vector similarities.

By providing the agent with a few-shot context consisting of relevant SQL queries, we provide the user of the system a way of telling the system what data is expected to be retrieved for different questions. This approach leverages the strengths of both retrieval and generation components, harnessing the domain knowledge encoded within the SQL query repository while enabling the agent to adapt and construct tailored queries based on the specific requirements of the user's input.

\subsection{Choosing an \gls{llm}} 

\glspl{llm} are used in the feature extractor and SQL agent components of the \frameworkname framework. In the \gls{llm} field, new open-source models such as LlamA 2~\cite{touvron2023llama,Llama-2} and Mistral-7B~\cite{jiang2023mistral} are rapidly emerging, but lack advanced function calling capabilities required by LangChain. While fine-tuning a model for feature extraction is viable, it demands substantial data and is time-consuming. Therefore, we decided on using OpenAI's \gptfour~\cite{OpenAIGPT4} and \gptthree~\cite{OpenAIGPT3} models.

\section{Evaluation}\label{section:evaluation}

\begin{table*}

    \scriptsize
    \centering
    \caption{Framework evaluation: extractor-validator evaluation (top), query complexity analysis (middle), ablation study (bottom).}
    \label{tab:evaluation-combo}
    
    \begin{tabular}{|c|c||c|c|c|c|c|c|c|c|c|c|}

        \hline
        \multicolumn{2}{|l||}{\textbf{Module Evaluation}}
        & \textbf{Q1} 
        & \textbf{Q2} 
        & \textbf{Q3} 
        & \textbf{Q4} 
        & \textbf{Q5} 
        & \textbf{Q6} 
        & \textbf{Q7} 
        & \textbf{Q8} 
        & \textbf{Q9} 
        & \textbf{Q10} 
        \\ \hline
        
        \multicolumn{2}{|l||}{Extractor} 
        & \medium 
        & \high 
        & \high 
        & \high 
        & \high 
        & \high 
        & \high 
        & \high 
        & \high 
        & \high 
        \\

        \multicolumn{2}{|l||}{Validator} 
        & \high 
        & \high 
        & \high 
        & \high 
        & \high 
        & \high 
        & \high 
        & \high 
        & \high 
        & \high 
        \\

        \multicolumn{2}{|l||}{Overall} 
        & \high 
        & \high 
        & \high 
        & \high 
        & \high 
        & \high 
        & \high 
        & \high 
        & \high 
        & \high 
        \\ 


        \hline
        \multicolumn{2}{|l||}{}
        & \textbf{Q11} 
        & \textbf{Q12} 
        & \textbf{Q13} 
        & \textbf{Q14} 
        & \textbf{Q15} 
        & \textbf{Q16} 
        & \textbf{Q17} 
        & \textbf{Q18} 
        & \textbf{Q19} 
        & \textbf{Q20} 
        \\ \hline
        
        \multicolumn{2}{|l||}{Extractor} 
        & \high 
        & \high 
        & \high 
        & \high 
        & \low\xspace / \high 
        & \high 
        & \high 
        & \medium 
        & \high 
        & \high 
        \\

        \multicolumn{2}{|l||}{Validator} 
        & \high 
        & \high 
        & \high 
        & \high 
        & \high\xspace / \high 
        & \high 
        & \high 
        & \high 
        & \high 
        & \high 
        \\

        \multicolumn{2}{|l||}{Overall} 
        & \high 
        & \high 
        & \high 
        & \high 
        & \medium\xspace / \high 
        & \high 
        & \high 
        & \high 
        & \high 
        & \high 
        \\ 
        
        \hline
        \hline

    
        \hline
        \multicolumn{2}{|l||}{\textbf{Query Complexity Metric}} 
        & \textbf{Q1} 
        & \textbf{Q2} 
        & \textbf{Q3} 
        & \textbf{Q4} 
        & \textbf{Q5} 
        & \textbf{Q6} 
        & \textbf{Q7} 
        & \textbf{Q8} 
        & \textbf{Q9} 
        & \textbf{Q10} 
        \\ \hline
        
        \multicolumn{2}{|l||}{Number of distinct operators, $n_1$} 
        & 5 
        & 6 
        & 8 
        & 8 
        & 8 
        & 10 
        & 10 
        & 9 
        & 8 
        & 9 
        \\
        
        \multicolumn{2}{|l||}{Number of distinct operands, $n_2$}  
        & 5 
        & 14 
        & 31 
        & 26 
        & 26 
        & 31 
        & 20 
        & 36 
        & 28 
        & 25 
        \\
        
        \multicolumn{2}{|l||}{Total number of operators, $N_1$} 
        & 5 
        & 11 
        & 33 
        & 18 
        & 18 
        & 22 
        & 16 
        & 27 
        & 19 
        & 19 
        \\
        
        \multicolumn{2}{|l||}{Total number of operands, $N_2$} 
        & 5 
        & 16 
        & 110 
        & 26 
        & 26 
        & 38 
        & 22 
        & 41 
        & 31 
        & 26 
        \\
        
        \multicolumn{2}{|l||}{Vocabulary, $n$} 
        & 10 
        & 20 
        & 39 
        & 34 
        & 34 
        & 41 
        & 30 
        & 45 
        & 36 
        & 34 
        \\
        
        \multicolumn{2}{|l||}{Length, $N$} 
        & 10 
        & 27 
        & 143 
        & 44 
        & 44 
        & 60 
        & 38 
        & 68 
        & 50 
        & 45 
        \\
        
        \multicolumn{2}{|l||}{Volume, $V$} 
        & 33.22 
        & 116.69 
        & 755.81 
        & 223.85 
        & 223.85 
        & 321.45 
        & 186.46 
        & 373.45 
        & 258.50 
        & 228.94 
        \\
        
        \multicolumn{2}{|l||}{Difficulty, $D$} 
        & 2.50 
        & 3.43 
        & 14.19 
        & 4.00 
        & 4.00 
        & 6.13 
        & 5.50 
        & 5.12 
        & 4.43 
        & 4.68 
        \\
        
        \multicolumn{2}{|l||}{Effort, $E$} 
        & 83.05 
        & 400.09 
        & 10727.66 
        & 895.39 
        & 895.39 
        & 1970.20 
        & 1025.54 
        & 1913.91 
        & 1144.77 
        & 1071.42 
        \\
        
        \multicolumn{2}{|l||}{Time to understand (s), $T$} 
        & 4.61 
        & 22.23 
        & 595.98 
        & 49.74 
        & 49.74 
        & 109.46 
        & 56.97 
        & 106.33 
        & 63.60 
        & 59.52 
        \\ 
        
        \hline
        \hline


        \hline
        \textbf{\gls{llm}}
        & \textbf{Pipeline}
        & \textbf{Q1} 
        & \textbf{Q2} 
        & \textbf{Q3} 
        & \textbf{Q4} 
        & \textbf{Q5} 
        & \textbf{Q6} 
        & \textbf{Q7} 
        & \textbf{Q8} 
        & \textbf{Q9} 
        & \textbf{Q10} 
        \\ \hline
        
        \multirow{6}{*}{\gptthree}
        & 1
        & \no 
        & \no 
        & \no 
        & \no 
        & \no 
        & \no 
        & \no 
        & \no 
        & \no 
        & \no 
        \\

        & 2
        & \yes 
        & \no 
        & \no 
        & \no 
        & \yes 
        & \no 
        & \no 
        & \no 
        & \no 
        & \no 
        \\
        
        & 3
        & \no
        & \no 
        & \no 
        & \no 
        & \no 
        & \no 
        & \no
        & \no 
        & \no 
        & \no
        \\

        & 4
        & \no 
        & \no 
        & \no 
        & \no 
        & \no 
        & \no 
        & \no 
        & \no 
        & \yes 
        & \no 
        \\
        
        & 5
        & \no
        & \yes 
        & \yes 
        & \no 
        & \no 
        & \yes 
        & \no 
        & \no
        & \no 
        & \no 
        \\

        & 6
        & \yes  
        & \yes 
        & \yes 
        & \yes 
        & \yes 
        & \yes 
        & \yes 
        & \no 
        & \yes 
        & \yes 
        \\ \hline
        
        \multirow{6}{*}{\gptfour}
        & 1
        & \yes 
        & \no 
        & \no  
        & \no 
        & \no 
        & \no 
        & \yes 
        & \no
        & \no 
        & \no
        \\
        
        & 2
        & \yes 
        & \no 
        & \yes  
        & \no 
        & \no 
        & \no 
        & \no 
        & \no
        & \yes 
        & \yes
        \\

        & 3
        & \yes 
        & \yes 
        & \no  
        & \no 
        & \no 
        & \no 
        & \yes 
        & \no
        & \no 
        & \no
        \\

        & 4
        & \yes  
        & \no 
        & \no 
        & \no 
        & \no 
        & \no 
        & \no 
        & \no 
        & \yes 
        & \yes 
        \\
        
        & 5
        & \yes 
        & \no 
        & \yes  
        & \no 
        & \yes 
        & \yes 
        & \yes 
        & \no
        & \no 
        & \yes 
        \\
        
        & 6
        & \yes  
        & \yes 
        & \yes 
        & \yes 
        & \yes 
        & \yes 
        & \no 
        & \no 
        & \yes 
        & \yes
        \\ 
        
        \hline

    \end{tabular}
    
\end{table*}

We use the following set of sample questions throughout our experiments to evaluate \frameworkname and its components\footnote{List available in~\cite{SoccerRAG-GitHub}, an extended set can also be found in~\cite{strand2024soccerrag}.}: \\

\begin{scriptsize}
    \begin{itemize}
        \item \textbf{Question 1:} Is Manchester United in the database? 
        \item \textbf{Question 2:} Give me the total home goals for Bayern M in the 2014-15 season.
        \item \textbf{Question 3:} Calculate home advantage for Real Madrid in the 2015-16 season
        \item \textbf{Question 4:} How many goals did Messi score in the 15-16 season?
        \item \textbf{Question 5:} How many yellow-cards did Enzo Perez get in the 15-2016 season?
        \item \textbf{Question 6:} List all teams that played a game against Napoli in 2016-17 season in seriea? Do not limit the number of results
        \item \textbf{Question 7:} Give all the teams in the league ucl in the 2015-2016 season?
        \item \textbf{Question 8:} Give me all games in epl with yellow cards in the first half in the 2015-2016 season
        \item \textbf{Question 9:} What teams and leagues has Adnan Januzaj play in?
        \item \textbf{Question 10:} List ALL players that started a game for Las Palmas in the 2016-2017 season? Do NOT limit the number of results .
        \item \textbf{Question 11:} Did Ajax or Manchester United win the most games in the 2014-15 season?
        \item \textbf{Question 12:} How many yellow and red cards were given in the UEFA Champions League in the 2015-2016 season?
        \item \textbf{Question 13:} Are Messi and C. Ronaldo in the database?
        \item \textbf{Question 14:} How many goals did E. Hazard score in the game between Bournemouth and Chelsea in the 2015-2016 season?
        \item \textbf{Question 15:} How many yellow cards were given in the game between Bayern Munich and Shakhtar Donetsk in the 2014-15 UEFA Champions League, and did anyone receive a red card?
        \item \textbf{Question 16:} Make a list of when corners happened in the English Premier League (EPL) 2015-2016 season. Aggregate by a period of 15 minutes.
        \item \textbf{Question 17:} What league is Manchester United, Arsenal, Bournemouth, Real Madrid, Chelsea, and Liverpool in?
        \item \textbf{Question 18:} How many players have "Aleksandar" as their first name in the database, and how many goals have they scored in total?
        \item \textbf{Question 19:} What did the commentary say about the game between Arsenal and Southampton in the 2016-17 season?
        \item \textbf{Question 20:} Have Mesut Ozil, Pablo Insua, or Alex Pike played for West Ham or Barcelona?
    \end{itemize}
\end{scriptsize}

\subsection{Extractor-Validator Evaluation}

The extractor-validator chain was evaluated on questions 1-20 subjectively, with the following evaluation criteria: (1) Were all properties extracted? (2) Were extra properties extracted? (3) Could extracted properties be validated? (4) Did the chain need human input? 
A perfect score (indicated with \high in Table~\ref{tab:evaluation-combo}) was given if all properties were extracted and validated. If any properties were missed or validation failed, the result was marked as failed (\low). Extracting extra information or needing human feedback resulted in a 50\% score (\medium).

\gptthree was used for all questions except 15 and 18, which were re-run with \gptfour. Table~\ref{tab:evaluation-combo} shows optimal performance for questions 2-10 and most of 11-20. In question 1, an extra item was extracted, without disrupting the rest of the chain. \gptthree failed to extract one property in question 15, which \gptfour handled correctly, this is demonstrated by presenting two results for Question 15. In Table~\ref{tab:evaluation-combo}, Question 18 was a question designed to confuse the extractor by asking about players named "Aleksandar". The extractor extracts the name, and then tries to validate it against a \textit{single} player. This shows that the chain does not understand context, as opposed to the SQL agent which is more content aware.

\subsection{Query Complexity Analysis}\label{sec:query-complexity-analysis}

To investigate the difficulty of composing database queries to represent natural language queries, query complexity scores were calculated for questions 1-10 using the Halstead metrics~\cite{Halstead}, which are frequently used to quantify the complexity of code snippets\footnote{Halstead metrics are a set of software complexity measures based on counting the distinct operators and operands in source code, used to estimate code complexity, maintainability, and the effort required for development and maintenance.}. For each of the 10 questions, a database query was manually composed, and the number of unique operators (e.g., "SELECT", "FROM", "=") and operands (e.g., table and column names) were counted, as well as the total occurrences of operators and operands across the query. Halstead metrics were then calculated based on these counts. The results (presented in the middle part of Table~\ref{tab:evaluation-combo}) show that question 3 had the highest complexity score. Further analysis revealed that this was due to the query for question 3 performing several \textit{SUM} operations.

\subsection{Ablation Study}

By testing the framework components under different configurations, it is possible to quantify the contribution of each component to the final \frameworkname framework. This information is crucial for understanding the strengths and weaknesses of the various parts of the system and identifying areas for potential improvement. We performed an extensive ablation study using the following pipeline configurations, where (6) corresponds to the full \frameworkname framework.\\

\begin{scriptsize}
    \begin{enumerate}
        \item \textbf{SQL agent only}: Sending the prompt directly to the SQL agent without \gls{rag} 
        \item \textbf{SQL agent with \acrshort{rag}}: Sending the prompt directly to the SQL agent with \gls{rag}
        \item \textbf{Extractor only}: Sending the prompt with extracted values to the SQL agent
        \item \textbf{Extractor and SQL \acrshort{rag}}: Sending the prompt with extracted values to the SQL agent, the SQL agent uses \gls{rag} to get examples
        \item \textbf{Extractor and validator}: Sending the prompt with extracted and validated values to the SQL agent
        \item \textbf{Extractor, validator, and SQL \acrshort{rag}}: Sending the prompt with extracted and validated values to the SQL agent, the SQL agent uses \gls{rag} to get examples
    \end{enumerate}
\end{scriptsize}

For the ablation study, questions 1-10 were used. A correct answer resulted in a pass score (indicated with \yes \ in Table~\ref{tab:evaluation-combo}), while an incorrect or incomplete answer resulted in a fail score (\no). The answers generated by the prompts created during the query complexity analysis (Section~\ref{sec:query-complexity-analysis}) were used to determine the success rate. The experimental setup involved using both \gptthree and \gptfour models to gain insights into their respective performance and to investigate the difference in quality, speed and cost between the two models. For the SQL-\gls{rag}, the value of K was set to 2.

\begin{figure}

    \centering   
    \begin{tikzpicture}
    
        \begin{axis}[
            ybar, 
            bar width=4.5pt,
            enlargelimits=0.15,
            ymax=250,
            legend style={at={(0.5,-0.15)},
            anchor=north,legend columns=-1},
            ylabel={Total Execution Time (s)},
            symbolic x coords={1,2,3,4,5,Average},
            xtick=data,
            nodes near coords,
            nodes near coords align={vertical},
            x tick label style={rotate=25,anchor=east},
            every node near coord/.append style={rotate=90, anchor=west, font=\scriptsize} 
            ]
            \addplot [color=midnightblue, fill=midnightblue] coordinates {(1,113.03) (2,125.90) (3,111.10) (4,122.01) (5,127.14) (Average,119.84)};
            \addplot [color=midnightblue, fill=midnightblue, pattern=dots] coordinates {(1,105.15) (2,96.1) (3,84.03) (4,100.96) (5,102.64) (Average,97.78)};
            \addplot [color=royalpurple, fill=royalpurple] coordinates {(1,232.90) (2,193.62) (3,195.51) (4,195.99) (5,238.12) (Average,211.23)};
            \addplot [color=royalpurple, fill=royalpurple, pattern=dots] coordinates {(1,144.23) (2,133.59) (3,139.91) (4,149.17) (5,155.74) (Average,144.53)};
            \legend{v3.5 Peak,v3.5 Off-peak,v4 Peak,v4 Off-peak}
        \end{axis}
        
    \end{tikzpicture}
    
    \caption{Total execution time for questions 1-10 with \gptthree and \gptfour.}
    \label{fig:execution-time}
    
\end{figure}
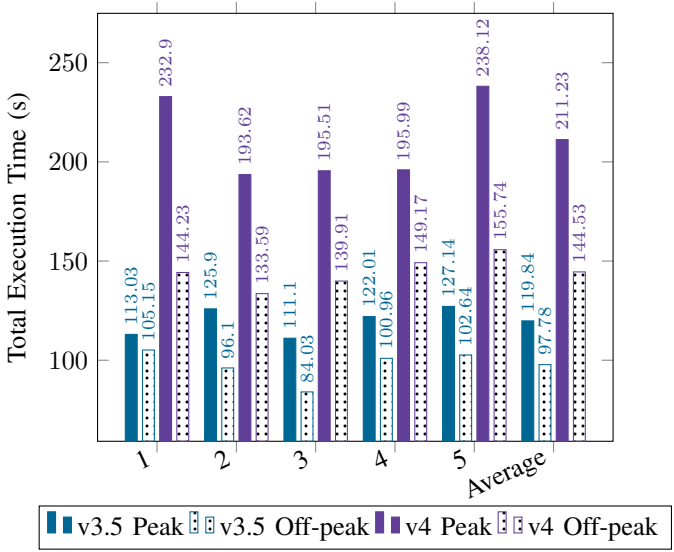

As shown in Table~\ref{tab:evaluation-combo}, the standalone SQL agent struggled with both models. The results indicate that the extractor-validator chain is crucial for increasing the hit rate. Interestingly, both models exhibited similar performance when the full pipeline was applied. \gptfour failed on question 7 due to model laziness, where it provided only a partial answer. For question 8, the query returned a substantial amount of text, causing the language model to refuse to provide a correct answer.

\subsection{Execution Time}

We investigated the execution time (total time from query until model output) for \gptthree and \gptfour. As the performance of cloud-based services might change based on global usage, we explored "peak" and "off-peak" periods separately (peak defined as 8AM-5PM EST and off-peak defined as 5PM-8AM EST). To observe and quantify the difference in execution time and response quality, we ran questions 1-10 consecutively, five times per batch, during peak and off-peak periods, with both models. Figure~\ref{fig:execution-time} presents our findings, which indicate that peak usage is related to working hours in the US, and that performance degrades during peak usage periods for both models, although \gptthree is significantly faster than \gptfour overall. 

\section{Discussion}\label{section:discussion}

\subsection{Handling Input Errors}

The extractor-validator chain has proven invaluable in mitigating spelling mistakes and other input errors and enhancing the overall accuracy of the system. Its functions can range from simple substitutions (e.g., heterogeneous season indicators in questions 5 and 6 into the uniform \textit{yyyy-yyyy} format, or "seriea" in question 6 to "Serie A", as expected by the database) to resolving more complex issues. In the examples below, we see that when the \gls{llm} doesn't manage to resolve keywords, it will in many cases produce a wrong output. 

\begin{scriptsize}
\begin{mdframed}
\noindent\textbf{Prompt:} Did Real or Man U win the most games in the 2015 season? \\
\noindent\textbf{Output:} Swansea won the most games in the 2014-2015 season with a total of 37 wins. \\
\noindent\textbf{Cleaned prompt:} Did Real or Man U win the most games in the 2015 season? \\
Updated Information:
\begin{itemize}
    \item Real (now referred to as Real Madrid) has a primary key: 30.
    \item Man U (now referred to as Manchester United) has a primary key: 7.
    \item 2015 (now referred to as the 2015-2016 season).
\end{itemize}
\noindent\textbf{Output:} In the 2015-2016 season, Real Madrid won 26 games, while Manchester United won 5 games. Therefore, Real Madrid won the most games in that season. 
\end{mdframed}
\end{scriptsize}

\subsection{Retrieving Statistics}

It is possible to extract detailed statistics using natural language queries with \frameworkname, as demonstrated below. 

\begin{scriptsize}
\begin{mdframed}
\noindent\textbf{Prompt:} Calculate home advantage for Real Madrid in the 2015-16 season \\
\noindent\textbf{Output:} For Real Madrid in the 2015-2016 season, the home advantage can be calculated as follows:
\begin{itemize}
    \item Total home goals scored: 55
    \item Total away goals scored: 39
    \item Number of home games played: 18
    \item Number of away games played: 17
\end{itemize} 
This indicates that Real Madrid scored significantly more goals at home compared to away games, suggesting a strong home advantage during the 2015-2016 season. 
\end{mdframed}
\end{scriptsize}

However, \glspl{llm} make mistakes with complex mathematics or large lists. We found that asking for a list often led to only a subset being returned (e.g., question 8, Prompt A), whereas asking for a count of events provided the correct answer (Prompt B=modified version of question 8). 

\begin{scriptsize}
\begin{mdframed}
    \noindent\textbf{Prompt A:} Give all the teams in the league ucl in the 2015-2016 season \\
    \noindent\textbf{Output:} The teams in the UEFA Champions League (UCL) during the 2015-2016 season, limited to five examples, are:
    \begin{enumerate}
        \item Galatasaray
        \item Atletico Madrid
        \item Manchester City
        \item Juventus
        \item Paris Saint-Germain (Paris SG)
    \end{enumerate}  
    \noindent\textbf{Prompt B:} Count all games in epl with yellow cards in the first half in the 2015-2016 season \\
    \noindent\textbf{Output:} There were 70 games in the English Premier League (england\_epl) with yellow cards in the first half in the 2015-2016 season. 
\end{mdframed}
\end{scriptsize}

\subsection{Influence of LLM}

\textbf{Execution time:} We discovered that the OpenAI API works slower during peak periods, which follow the US timezone, with the execution time 22.56\% slower for \gptthree, and 46.5\% slower for \gptfour.

\textbf{Robustness:} Robustness refers to providing the same correct answer when presented with the same question multiple times. We did not observe significant degradation in the robustness of responses during peak periods. However, responses can exhibit variations across different runs, even for the same input. While strict prompting techniques can mitigate this inconsistency to some extent, the inherent stochastic nature of \glspl{llm} poses challenges in achieving complete determinism. To address this issue and gain greater control over the output, a solution would be to explore the usage of self-hosted \glspl{llm} for the framework. Self-hosted \glspl{llm} also provide users with control over the underlying network conditions, computational load, request prioritization, etc. that might influence the quality of the response.

\textbf{Model laziness:} The phenomenon where an \gls{llm} prematurely stops a task without fully completing it~\cite{Wei2024Feb} can be manifested in two ways. \textit{Model stopping before querying:} While attempting to answer a question, the framework makes multiple requests to the \gls{llm}. We observed instances where the framework prematurely halted during this reasoning phase, responding to the user with "I now have the information I need, and should query the database for the answer." This was addressed by including "I will not stop until I query the database and return the answer." in the system prompt for the \gptthree model, which seemed to fix most issues. The \gptfour model did not exhibit this problem. 
(2) \textit{Model returning a subset:} Both models faced this issue when asked to return a list. Despite fetching the correct data, the \glspl{llm} would often limit their response to 5-10 items. Stricter prompting did not resolve this issue, which is acknowledged by OpenAI~\cite{Coldewey2024Jan}.

\subsection{Future Work}

\frameworkname supports multimodal data integration by connecting metadata about game events with video timestamps, enabling several future applications. These include searching and retrieving specific video clips associated with particular game events using metadata, performing tasks such as object detection, action recognition, and scene understanding on video clips, and ultimately, automatically generating video highlights centered around game events of interest, or entire game summaries. \frameworkname also faces several limitations that need addressing. First, the system struggles with complex queries and large data volumes. \gls{llm} model laziness might cause tasks to conclude prematurely or provide partial data, especially with extensive outputs. There is also a need to explore emerging \glspl{llm}, including open-source alternatives, for potential performance improvements with respect to function calling, language to query translation, and extraction abilities. Multimodal data synchronization and real-time processing capabilities require significant improvement to be viable for live sports analytics applications. Finally, ensuring privacy and ethical data handling is paramount, necessitating robust security measures and ethical guidelines in future developments.

\section{Conclusion}\label{section:conclusion}

We introduce \frameworkname, a framework that leverages \gls{rag} and \glspl{llm} to efficiently retrieve multimodal soccer information via natural language queries, enhancing the accessibility of vast sports datasets by allowing for intuitive user interactions with complex data archives. Our evaluation demonstrates \frameworkname's capability to accurately interpret complex queries and facilitate dynamic user engagements with multimedia content. Ultimately, \frameworkname exemplifies the transformative potential of AI in sports analytics, promising a future where technology enriches the global fan experience and advances the accessibility of sports information.

\section*{Acknowledgment}

This research was partly funded by the Research Council of Norway, project number 346671 (AI-Storyteller). 

\clearpage
\balance
\bibliographystyle{IEEEtran}
\bibliography{references}

\end{document}